\documentclass[12pt]{article}

\usepackage{psfig}

\newcommand{\bibl}[5]
	{#1, {\it #2} {{\bf #3},} #5 (#4)}
\newcommand{\cebe}{\begin{center}}
\newcommand{\ceen}{\end{center}}
\newcommand{\debe}{\begin{description} \vspace{-2ex}}
\newcommand{\deen}{\end{description}}
\newcommand{\eabe}{\begin{eqnarray}}
\newcommand{\eaen}{\end{eqnarray}}

\newcommand{\eqbe}{\begin{equation}}
\newcommand{\eqen}{\end{equation}}

\newcommand{\flbe}{\begin{flushright} \vspace{-2ex}}
\newcommand{\flen}{\end{flushright}}

\newcommand{\itbe}{\begin{itemize}}
\newcommand{\iten}{\end{itemize}}

\newcommand{\tabe}{\begin{tabbing}}
\newcommand{\taen}{\end{tabbing}}

\newcommand{\epe}{\mbox{e}^-}
\newcommand{\epea}{\mbox{e}^+}

\newcommand{\epq}{\mbox{q}}

\newcommand{\epqa}{\overline{\mbox{q}}}

\newcommand{\ggn}{\mbox{g}}
\newcommand{\gwm}{\mbox{W}^-}
\newcommand{\gwp}{\mbox{W}^+}

\newcommand{\mpn}{\pi^0}

\newcommand{\quadq}{\epq_1\epqa_2\epq_3\epqa_4}

\begin{document}


\begin{titlepage}

\begin{flushright}
LU TP 97-32 \\
November 1997
\end{flushright}
\vspace{25mm}
\cebe
\Large
{\bf Bose-Einstein and Colour Interference \\ in W-pair Decays} \\
\normalsize
\vspace{12mm}
Jari H\"akkinen and Markus Ringn\'er\footnote{jari@thep.lu.se, markus@thep.lu.se} \vspace{1ex} \\
Department of Theoretical Physics, Lund University, \\
S\"olvegatan 14A, S-223 62 Lund, Sweden 
\ceen
\vspace{20mm}

\noindent {\bf Abstract} \\ We study effects on the W mass
measurements at LEP2 from non-perturbative interference effects in the
fully hadronic decay channel. Based on a model for Bose-Einstein
interference, which is in agreement with LEP1 data, we argue that
there are no Bose-Einstein correlations between bosons coming from the
different W's. For small reconnection probabilities we rule out the
possible experimental signal of colour interference at LEP2, suggested
in~\cite{r:gg94}. The conclusions from this paper are that the
theoretical uncertainties in the W mass determination should be
smaller than the experimental statistical error.
\end{titlepage}


\section{Introduction} 
One of the main goals of LEP2 is to perform high quality precision
measurements of the W mass. In order to obtain the projected
statistical error of 30--40~MeV, all decay channels -- the leptonic,
the semi-leptonic, and the hadronic -- have to be used. The purely
leptonic decays will however be rare and they will not have a large
impact on the measurements. In the two cases involving hadronic
systems non-perturbative effects, such as colour- and Bose-Einstein
interference, can occur and the measured W mass may be affected. The
interference effects within the hadronic system can in the
semi-leptonic case be estimated from LEP1 studies. From these studies
we understand the effects of Bose-Einstein (BE) correlations quite
well and we have also learnt that the colour interference (CI) effects
are probably small. This means that the semi-leptonic case can be
reconstructed using a Monte Carlo tuned to LEP1 data and that the
theoretical uncertainties due to interference effects will only
influence the fully hadronic channel. These uncertainties arise since
the interference effects may have impact on the identity of the two
decaying W's.  The fully hadronic channel is very nice since we can,
in principle, observe all the momentum of the event. However even if
LEP2 provides enough statistics for a sub 30~MeV error the
interference effects have to be taken into account, or at least be
under theoretical control.

That Bose-Einstein correlations might affect the measurement of the
mass of the W at LEP2 was first suggested in \cite{r:ls95}. The
typical separation in space and time between the $\gwp$ and
$\gwm$ decay vertices is smaller than 0.1 fm in fully hadronic
events, i.e. $\epea\epe \rightarrow \gwp\gwm \rightarrow
\quadq$, at LEP2 energies \cite{r:skz94,r:skpl94}.
Since this distance is much smaller than typical hadronic sizes and
the correlation lengths associated with Bose-Einstein effects, pions
from different W's are argued to be subject to Bose-Einstein
symmetrisation. The effect on the W mass has been estimated in a
number of models with widely varying results
\cite{r:ls95,r:eg96,r:beweights97}.  
In this paper we will based on the model in \cite{r:ar97}
argue that there are no Bose-Einstein correlations between particles
stemming from different W's at LEP2. We will also discuss the
consequences of the symmetrisation for various ways of reconstructing
the W mass.

Colour interference can occur in the W-pair decays at LEP2 but the
probability for reconnections is unknown. In this study we use an
improved Monte Carlo implementation of the model described
in~\cite{r:gg94} to address the possibility to experimentally detect
effects from CI at LEP2. We will also use it to estimate the effect of
CI on the W mass determination.

After a short description of the various mass reconstruction schemes
we use, we will in section~\ref{s:models} describe the important
features of our interference models. We will in particular review how
the correlation length in our BE model arises stressing the parts
relevant to understand correlations between particles from different
W's. This is followed by the results for the reconstruction of the
W mass and conclusions.

\section{Mass reconstruction} 
\label{s:reconstruction}

If every final particle in the fully hadronic case can be uniquely and
correctly assigned to either the $\gwp$ or the $\gwm$ decay, the
$\mbox{W}^{\pm}$ four-momenta can be reconstructed and squared to give
the $\mbox{W}^{\pm}$ masses. There are however many complications
which have to be taken into account in practice. It is not our
intention to cover these complications here, but a detailed discussion
in can be found in \cite{r:skz94} together with a discussion about
various ways to reconstruct the W mass in order to avoid
complications. Reconstruction schemes are devised in~\cite{r:skz94} to
study the effects of interference and we have adopted some of them in
the our analysis. We will only give a brief sketch of how it is done
and the reader is referred to the original work for details.

Four jet events are selected using the LUCLUS
algorithm \cite{r:JETSET}, with the jet distance parameter
$d_{join}=8~\mbox{GeV}$. This rejection of events with hard gluon jets is
done since they give a much worse W mass resolution. In addition, we
require the jets to have energies above 20~GeV and that the angle
between any two jets is greater than 0.5 radians, to reduce the number
of misassignments.  The four jets can be paired in three different
ways giving different results for the W mass. We use three different criteria 
to single out one combination.

\begin{description}
\item[1]: The pairs are chosen so that the deviation of 
the average reconstructed W mass from the used mass is minimized;
\[
\mbox{min}\left| \frac{M_{\mbox{\scriptsize W}^+} + M_{\mbox{\scriptsize W}^-}}{2} - M_{\mbox{\scriptsize W}} \right|.
\]
This is not measurable in an experimental situation since we cannot
know with which masses the W's were produced, but it is included for
comparison.

\item[2]: The pairs are chosen so that the deviation of 
the sum of the reconstructed masses from a known
nominal mass is minimized;
\[
\mbox{min}(\left| M_{\mbox{\scriptsize W}^+} - M_{\mbox{\scriptsize W}} \right| + \left| M_{\mbox{\scriptsize W}^-} - M_{\mbox{\scriptsize W}} \right|).
\]

\item[3]: The pairs are chosen so that the sum of their opening angles 
is maximized. This makes sense close to threshold where the jets from
the same W should be almost back-to-back.
\end{description}
To investigate the effects of the interference models we compare the
reconstructed W mass with interference with the
reconstructed mass without interference.


\section{Models}
\label{s:models}
Before going into the details of our models we will shortly discuss
some general features of $\mbox{W}\mbox{W} \rightarrow \quadq$
events, which provide a motivation for some of our assumptions.  As
will be made clear, our models for the interference effects and in
particular some of their major consequences are based upon the picture
of singlet strings fragmenting. This may however not be the
full story, since there could be an important non-singlet component of
hadronisation, especially in the scenario when two strings are formed
close to each other. The only hadronisation model which includes a
non-singlet component is that of Ellis and Geiger
\cite{r:eg96}. In the case of a non-singlet component in
$\mbox{W}\mbox{W}\rightarrow \quadq$ one would expect that the
multiplicity in W-pair events is different from twice the
multiplicity in single string events. This is manifested in particular
in the colour reconnection scheme of Ellis and Geiger, where not only
the W mass shift is much larger than in their singlet models, but
it also results in a substantial reduction of the number of hadrons coming
from the overlap region of the two W's.

Three of the LEP experiments (DELPHI/L3/OPAL) have measured the mean
charged hadronic multiplicity in $\gwp\gwm\rightarrow
\quadq$ events, $\langle N_{ch}^{4\mbox{\scriptsize q}} \rangle$, 
and in $\gwp\gwm\rightarrow\epq\epqa\mbox{l}{\overline{\nu}}_{\mbox{\scriptsize l}}$ events,
$\langle N_{ch}^{\mbox{\scriptsize qql}\nu}\rangle$
\cite{r:mult_delphi97,r:mult_l397,r:mult_opal97}. Summarizing their
results give \cite{r:ward97}
\eqbe 
\frac{\langle N^{4\mbox{\scriptsize q}}_{ch} \rangle}{2\langle
N^{\mbox{\scriptsize qql}\nu}_{ch} \rangle} = 1.04 \pm 0.03 
\label{e:exp_mult}
\eqen
which gives no support for models leading to a reduction of the
hadronic multiplicity in W-pair events.  This suggests that
singlet strings provide a good description of $\gwp\gwm\rightarrow
\quadq$ hadronisation.

\subsection{Colour interference at LEP2 energies}
The CI model in this paper is an improved Monte Carlo implementation
of the model described in~\cite{r:gg94}. The model for recoupling is
quite simple and its features are described in detail
in~\cite{r:gg94}. Here we give a summary of the model with emphasis
on the improvements.

The space--time distance between the W decay points in
$\epea\epe\rightarrow \gwp\gwm \rightarrow \quadq$ is
about $1/\Gamma_{\mbox{\scriptsize W}}$ and hard gluons with energies above $\Gamma_{\mbox{\scriptsize W}}$ are
therefore emitted incoherently by the two quark systems early in the
event \cite{r:yd93}. This means that there are two sets of partons
before any possible colour interference can occur. The two sets
$\epq_1\ggn_1\ggn_2\ldots \ggn_n \epqa_2$ and $\epq_3\ggn_{1^{'}} \ggn_{2^{'}}\ldots \ggn_{m^{'}}\epqa_4$ have a lot of different recoupling possibilities
since every set of particles $\epq\ldots \ggn$ is a
colour-triplet. Recoupling of a $\epq\ldots \ggn$ with any $\ggn\ldots \epqa$
from the other set can occur with the probability $1/N_{c}^2$ so the
total probability for recoupling can in principle be very large.
The estimation of the total recoupling probability is
non-trivial. In~\cite{r:gg94} a discussion is made about what kind of
probabilities to expect. No real conclusion was or can be
made, and the probability remains a free parameter of the model.

Perturbative QCD favours states which correspond to short strings
i.e. parton states which produce few hadrons. The $\lambda$ measure
was introduced in~\cite{r:ba88} and is a measure of the effective
rapidity range inside which the decay products of a particular
colour-singlet string are distributed.  In this way it is related to the
multiplicity. In \cite{r:gg94} it is argued that states with smaller
$\lambda$'s could be dynamically enhanced, and that this choice also
gives reconnected events that differ most from non-reconnected
systems. Reconnected states with the smallest $\lambda$ measure are
therefore chosen in the model.

All of this is still true in the CI model in this paper. We have
however made significant improvements in the MC implementation. The
Ariadne MC v4.08~\cite{r:ARIADNE} allows the user to stop the
production of gluons below some given energy. This feature was not
available in the original work, where gluons with energy below
$\Gamma_{\mbox{\scriptsize W}}$ where simply neglected (leading to a
3\% loss of energy). Furthermore, the W-pairs were incorrectly
generated in the original work since no spin information was preserved
and the W's were therefore allowed to decay isotropically. In order
to take the full angular correlations into account we now use Pythia
v5.7~\cite{r:JETSET}, where the full $2\rightarrow 2 \rightarrow 4$
matrix elements are included for the W-pair production and decay.

These improvements will lead to consequences for the results
obtained in~\cite{r:gg94}. In addition to studying possible
experimental signals at LEP2 of recoupled events we also extend the
analysis of~\cite{r:gg94} to study CI effects on W mass
determination.

\subsection{Bose-Einstein correlations in W-pair production} 

A model for Bose-Einstein correlations based upon a possible
quantum-mechanical framework for the Lund Fragmentation Model~\cite{r:ba83} 
has been proposed \cite{r:ah86} and it has been extended to the multi-particle
correlations needed at LEP energies \cite{r:ar97}.  An important
feature of the model is that it can be used as an extension of the
probability based Lund Model, implementing the correlations as event weights.

The interpretation of the Lund Fragmentation Model in \cite{r:ar97}
gives an explicit form for the transition matrix element for a string
fragmenting into hadrons. The resulting matrix element depends only on
the space-time history of the string and the model therefore uniquely
predicts the relative amplitudes for different particle
configurations, and therefore also the magnitude of the Bose-Einstein
effect. To understand how the correlation length between pions arise
in the model we will in the following shortly discuss the basic steps
leading to the specific form of the transition matrix element.

A unique breakup rule for a string can be derived inside the Lund Model,
which results in the following probability for a 
string to decay into hadrons $(p_1,\dots,p_n)$,  
\eqbe
dP(p_1,...,p_n) = \left[\prod_i (N dp_i \delta(p_i^2-m_i^2))\right] \delta(\sum_j p_j - P_{tot
})\exp(-bA)
\label{e:prob}
\eqen
where $A$ is the space-time area spanned by the string during its
break-up into $\epq\epqa$-pairs, and $N$ and $b$ are two free parameters.

The production of hadrons from a single string in the Lund Model 
can be given a quantum mechanical 
interpretation inside a non-Abelian field theory. The transition 
matrix element, ${\cal M}$, can, to obtain the result in Eq~(\ref{e:prob})
be identified with (note the similarity with Fermi's golden rule)
\eqbe
{\cal M} =  \exp(i\xi A) ~~\mbox{with}~~ \xi = \frac{1}{2\kappa}+\frac{ib}{2} 
\label{e:lmM}
\eqen
where the decay surface area, $A$, is in energy-momentum units in the
light-cone metric.  The imaginary part of the quantity $\xi$ is
related to the pair production probability.  As discussed in
\cite{r:ar97} the phase for ${\cal M}$, as given by the real part of
$\xi$, is found by observing how gauge invariance will constrain the
production of $\epq\epqa$-pairs along the colour force fields. The
main observation is that a final state hadron stems from a $\epq$ from
one vertex and a $\epqa$ from the adjoining vertex. This implies that
in order to keep gauge-invariance it is necessary that the production
matrix element contains at least a gauge connector,
$\exp{(ig\int^{j+1}_{j}A^{\mu}dx_{\mu})}$, between the two vertices,
denoted $j$ and $j+1$.  The total matrix amplitude for a single string
must contain at least one gauge connector for each hadron and we get a
Wilson Loop Operator as a minimal requirement for gauge invariance
\eqbe
{\cal M} = \exp(ig\oint A_{\mu}dx^{\mu})
\eqen
where the integral is around the decay surface of the string. 
Using Wilson's confinement criteria for the
behaviour of such a loop operator we get the real part of $\xi$, as in
Eq~(\ref{e:lmM}).  

The transverse momentum generation will also contribute to the total
matrix element. This contribution is discussed in detail in
\cite{r:ar97} and is found to be
\eqbe
\propto \exp(-\frac{1}{4\sigma ^2}{{\bf k}}^{2}_{\perp})
\label{e:lmptM} 
\eqen
where $\pm{\bf k}_{\perp}$ are the compensating transverse momenta 
generated in a $\epq\epqa$-vertex and $\sigma$ is the width of the Gaussian 
supression of the quarks transverse momenta.

\begin{figure}[t]
  \hbox{\vbox{
    \begin{center}
    \mbox{\psfig{figure=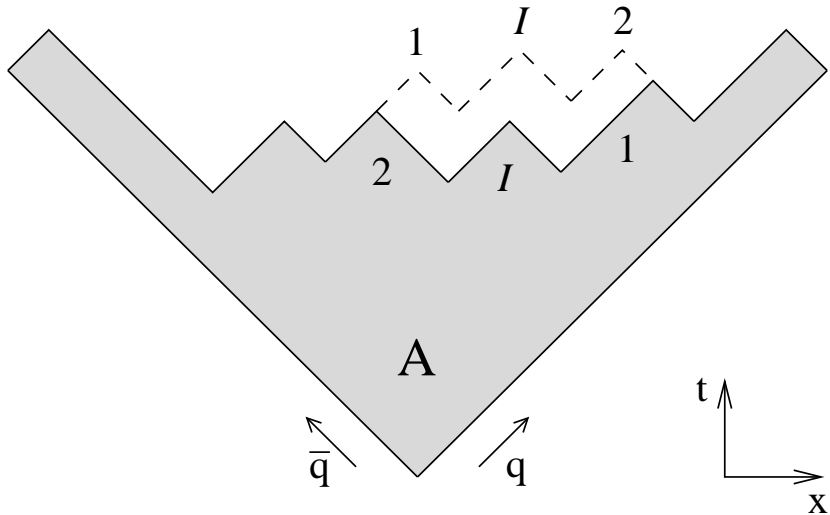,width=11.0cm}}
    \end{center}
  }}
\caption{\em The two possible ways, $(1,I,2)$ and $(2,I,1)$, to produce the entire state when $1$ and $2$ are identical bosons. The space-time area, $A$, spanned by the string during its break-up is shaded.}
\label{f:areadiff}
\end{figure}

In order to see the main mechanism for BE-correlations in the Lund
Model we consider Fig~\ref{f:areadiff}, in which two of the produced hadrons, 
denoted $(1,2)$, are assumed to be identical bosons and the state in between
them is denoted $I$. There are two ways to produce the entire state,
corresponding to exchange of the two identical bosons. The two configurations,
 $(\ldots,1,I,2,\ldots)$ and $(\ldots,2,I,1,\ldots)$, are shown in the figure 
and in general they correspond to different areas $A$.

The area difference, $\Delta A$, depends not only on the energy momentum 
vectors $p_1$ and $p_2$, but also on the four-momentum of the intermediate
state, $p_I$. The difference can be written as
\eqbe
\frac{\Delta A}{2\kappa} = \delta p \delta x
\label{e:deltaA}
\eqen
where $\delta p = p_2-p_1$ and $\delta x = (\delta t ,0,0,\delta z)$
is a reasonable estimate of the space--time difference, along the
string surface, between the production points. This means that the
correlation length, which is being measured by the four-momentum
difference between pairs, is in the model dynamically implemented as
$\delta x$ \cite{r:ar97}.  The correlation length is therefore
not the direct distance between production points. Instead it is the
distance along the string surface, i.e. the distance along the colour
force field. This is not surprising if we consider how the
quantum-mechanical process corresponding to the Lund Model was
derived; to keep gauge-invariance we got a gauge-connector between
adjacent vertices and this is what provides us with the $A/(2\kappa)$
factor in the matrix element, from which the correlation length in the
model stems.

In the case of production of two strings, i.e. a $\quadq$ system,
there is no reason for a gauge-connector between vertices belonging to
different strings. We will therefore assume that the distance along
the gauge-field between them is infinite even though the direct
space--time distance may be very small. This implies that there is no
interference between production vertices belonging to different
strings. This means that in this model each string can be considered a
system of its own, with separate Bose-Einstein effects. The resulting
event weight is then of course the product of the weights for each
system separately. In~\cite{r:ar97} it is explained how the BE
interference can be incorporated in a probabilistic event generation
scheme by weighting the produced events. In particular using the amplitudes
Eq~(\ref{e:lmM}) and Eq~(\ref{e:lmptM}) results in the weight
\eqbe
 w = \prod_{n=1}^2 \left(1+\sum_{{\cal P}^{\prime}_n\neq {\cal P}_n}\frac{\displaystyle \cos \frac{\textstyle \Delta A_n}{\textstyle 2\kappa}}
{\displaystyle \cosh\left( \frac{\textstyle b\Delta A_n}{2}+\frac{\textstyle \Delta (\sum^{(n)} p^{2}_{\perp q})}{\textstyle 2\sigma^{2}_{p_\perp}}\right)}\right)
\label{e:WWweight}
\eqen
for a fully hadronic WW event, where $ \Delta $ denotes the difference
with respect to configurations ${\cal P}_n $ and ${\cal P}^{\prime}_n
$ of the string $n$ and the sum of $p^{2}_{\perp q}$ is over all the
vertices of string $n$.  We have introduced $\sigma_{p_\perp}$ as the
width of the transverse momenta for the generated hadrons, (i.e.
$\sigma^{2}_{p_\perp}= 2\sigma^2$).

It should be emphasized that if only colour-singlet combinations of
partons are allowed to be formed there is no model consistent way to
get correlations between particles stemming from the different W's.
In comparison to most of the models using event weights to implement
BE-correlations \cite{r:beweights97} we have a physical picture of how
the correlation length in our model arises and it describes data well
in single string fragmentation \cite{r:ar97}.  Taken together with our
previous discussion of multiplicities in WW events this supports our
conclusion that there are no correlations between particles from the
different W's.


\section{Results}

All the results are for W-pairs generated at 170~GeV. We have checked
the effects of our models on the mean charged multiplicity and the
results are shown in Table~\ref{t:mult-shifts}.
\begin{table}[ht]
\begin{center}
\begin{tabular}{|cc|c|c|}
\hline
Model & & $\langle N^{4q}_{ch}\rangle$ & $\Delta\langle N^{4q}_{ch}\rangle~(\%) $ \\
\hline
CI              & without & 38.62 $\pm$0.01 & \\ 
                & 100\%   & 36.90 $\pm$0.01 & \\
                & 10\%    & 38.45 $\pm$0.01 & -0.44 \\
\hline
BE              & without & 24.4            &   \\
                & with    & 25.1            & +2.7$\pm$0.3  \\
\hline
\end{tabular}
\end{center}
\caption{\em The mean charged multiplicity for the two interference models. For the CI model we show the results for 100\% recoupled events and for an admixture of recoupled and non-recoupled events of the order 10\%. The lower multiplicities for the BE results are due to that no parton cascade has been used in this case.}
\label{t:mult-shifts}
\end{table}
We get small effects on the mean multiplicity and they are compatible
with the experimental result, Eq~(\ref{e:exp_mult}).

\subsection{Colour interference results}
It is natural to divide the CI results into two independent
parts. First we discuss the possibility to detect signals of CI at
LEP2 in the same way as it was done in~\cite{r:gg94} and then we will
study mass reconstruction effects.

\subsubsection{CI signal search at LEP2}
\begin{figure}[t]
\hbox{\vbox{
	\begin{center}
	\mbox{\psfig{figure=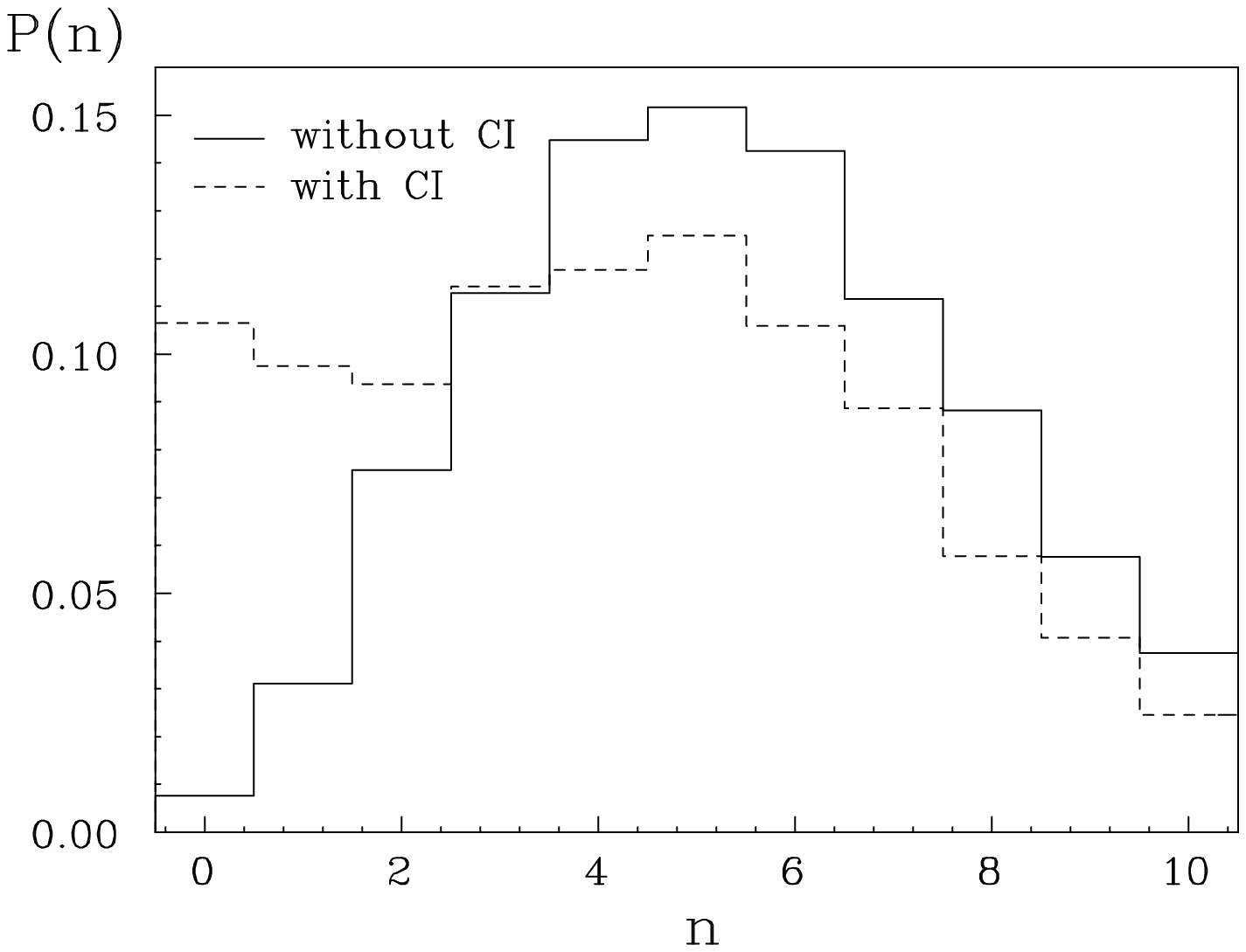,width=8.1cm}}
	\mbox{\psfig{figure=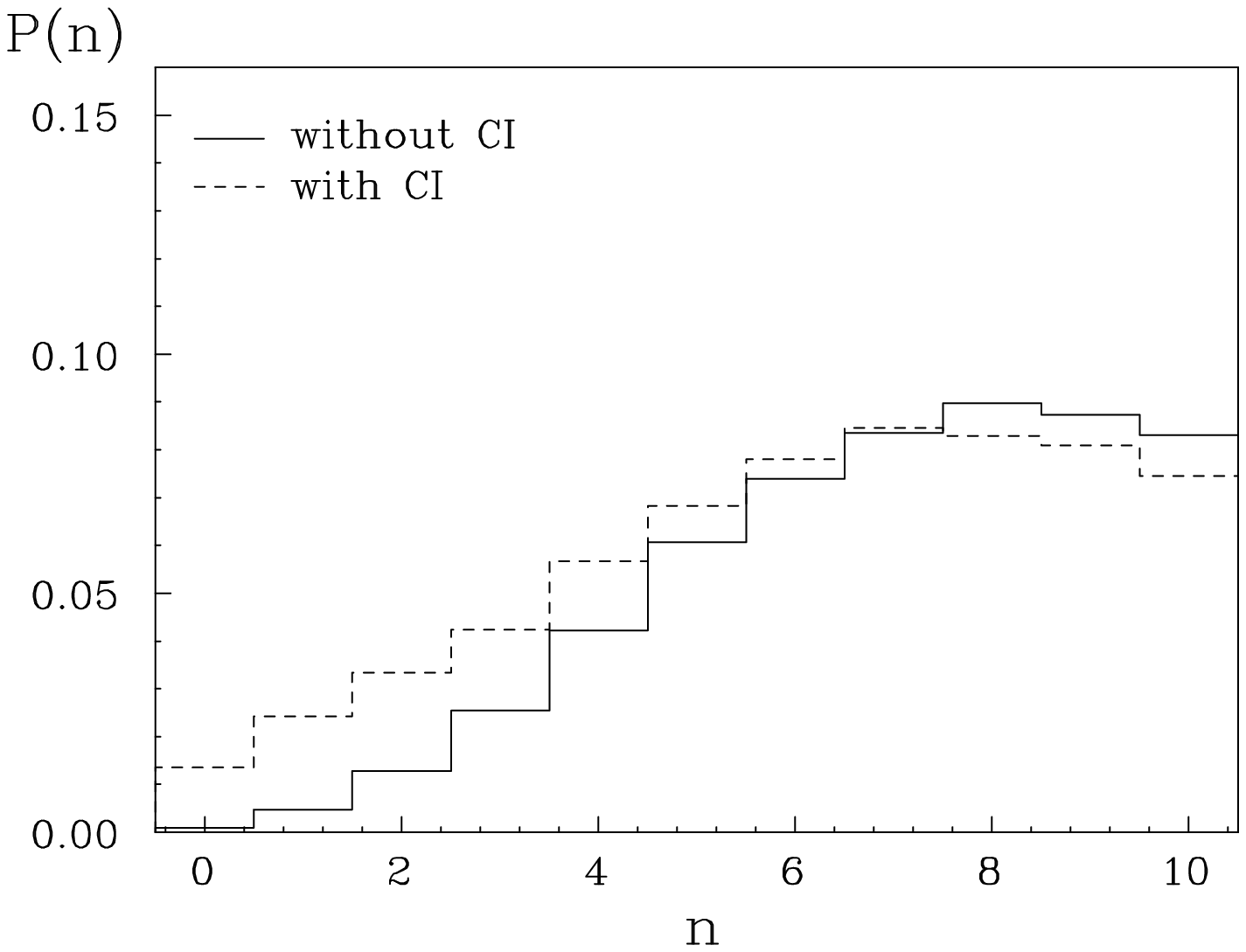,width=8.1cm}}
	\end{center}
}}
\caption{\em Multiplicity distributions for $\left|y\right|<0.5$ for non-recoupled (solid line) and recoupled events (dashed line) with thrust cuts \em left)  \em $T>0.92$ and \em right) \em $T>0.76$. The CI results are for 100\% recoupled events.}
\label{f:gapevents}
\end{figure}

The search for signals of CI at LEP2 in~\cite{r:gg94} give numbers
which are very close to what will be statistically significant with
the expected number of events from LEP2, when a 10\% recoupling
probability is assumed. The improvements made here will dilute the
signal proposed in \cite{r:gg94}. Multiplicity distributions
(including $\mpn$) in the central rapidity region,
$\left|y\right|<0.5$, for recoupled and non-recoupled events are shown
in Fig~\ref{f:gapevents}. Comparing these with the original results
from~\cite{r:gg94} we note that the signal-to-background ratio is
significantly reduced. 
\begin{table}[ht]
\begin{center}
\begin{tabular}{|c|c|c|c|c|}
\hline
Model & Thrust & Event fraction & Events with $n_{central}$=0 & background \\
\hline
\cite{r:gg94} & 0.92 & 0.04 & 4.3 & 0.68 \\
& 0.76 & 0.60 & 13 & 1.9 \\
\hline
our & 0.92 & 0.01 & 0.93 & 0.36 \\
& 0.76 & 0.60 & 6.4 & 2.6 \\
\hline
\end{tabular}
\end{center}
\caption{\em Expected number of events with zero particles in a central rapidity region: $\left|y\right|<0.5$, denoted by $n_{central}$, for a total of 5000 fully hadronic W-pair events. A 10\% recoupling probability is assumed.}
\label{t:gapevenets}
\end{table}
In Table~\ref{t:gapevenets} we have compiled the number of events
without particles in a central rapidity region at LEP2 using two
different thrust cuts and an expected 5000 fully hadronic events. We
note that the signal decreases and if this signal is to be seen at
LEP2 there must be a larger recoupling probability than 10\%. A larger
recoupling probability would increase the number of events without
particles in the central rapidity bin. A closer examination of the
improvements of the MC implementation in this paper reveals that the
conservation of energy will not change the result too much from the
original work. Almost all of the suppression of the signal comes from
taking the anisotropy of the W decays into account.

From this study we conclude that the statistics from LEP2 will make it
hard to use the signal proposed in~\cite{r:gg94}.

\subsubsection{W mass reconstruction results}
\label{sss:ciwmass}
We have studied the effects of CI on the W mass measurement to estimate
the size of the theoretical error on the mass implied by our model.

\begin{figure}[t]
\hbox{\vbox{
	\begin{center}
	\mbox{\psfig{figure=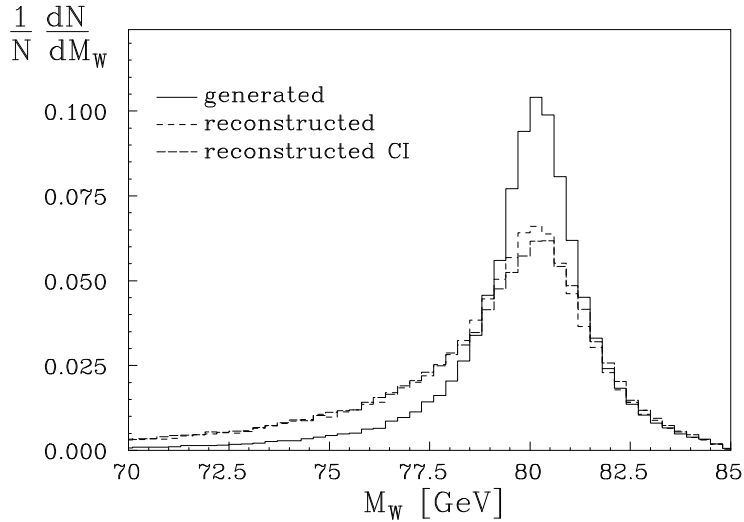,width=11cm}}
	\end{center} }}
\caption{\em The distribution of the  generated W mass (dashed)
 together with the reconstructed mass with (dotted) and without (solid) colour
 interference. The results are for reconstruction method~2.}
\label{f:cimass}
\end{figure}

In Fig~\ref{f:cimass} we show the generated W mass and the
reconstructed masses with and without CI interference. We see that the
difference between the reconstructed distributions is small.
The mass shifts for 100\% reconnected events are shown in
Table~\ref{t:cimass} and if the reconnection probability is assumed to
be 10\% the shifts should be scaled down with a factor 10. $\Delta M$
denotes the mass shift due to the reconstruction method as compared
with the generated W mass and the additional shift due to the
interference is denoted by $\delta M$.
\begin{table}[ht]
\cebe
\begin{tabular}{|c|r|r|}
\hline
Method & $\Delta M$ [MeV] & $\delta M$ [MeV] \\
\hline
1 & -279 $\pm$ 15 & -3 $\pm$ 21 \\ 
2 & -1238 $\pm$ 19 & -90 $\pm$ 27 \\ 
3 & -75 $\pm$ 16 & -27 $\pm$ 23 \\ 
\hline
\end{tabular}
\ceen
\caption{\em Shifts in the reconstructed W masses using the different methods from Section~\ref{s:reconstruction}.  $\Delta M$ denotes the
mass shift due to the reconstruction method and $\delta M$ denotes the additional shift due to the colour interference.}
\label{t:cimass}
\end{table}

Assuming a 10\% reconnection probability the shifts will be small and
negligible from the experimental mass reconstruction point of view.
However, in a worse case scenario with a 100\% probability the shifts
can be quite large but the experimental signal suggested
in~\cite{r:gg94} would then on the other hand be observable.

\subsection{Bose-Einstein interference results}
We have studied how the inclusion of Bose-Einstein correlations,
implemented as event weights, affect the results from various mass
reconstruction schemes. The main concern in \cite{r:ls95} was that the
BE effects in the hadronisation stage can couple identical particles
from the $\gwp$ and the $\gwm$. They used the LUBOEI algorithm
\cite{r:ls95} in which the momenta of the produced bosons are
reshuffled to reproduce a chosen BE-correlation. The momenta are then
rescaled by a common factor to keep energy-momentum conservation for
the event as a whole. This procedure might result in a redistribution
of momenta in such a way that the hadrons which come from the $\gwp
(\gwm)$ decay don't add up to the same invariant mass as the original
$\gwp(\gwm)$ had. It should be noted that the rescaling procedure
needed afterwards introduces shifts in the W mass even if there are no
BE-correlations between particles stemming from different W's. After
corrections for this 'spurious' mass shift a shift of about +100~MeV
at 170 c.m. energy was found in \cite{r:ls95}.

The main feature of our model is that we don't expect a coupling
between particles coming from different W's. The inclusion of
correlations may however affect for example multiplicities and event
shape variables and therefore it may affect the reconstruction of the
W boson mass. Such an artificial mass shift is hoped to be taken into
account by the tuning of the JETSET MC~\cite{r:JETSET} to the
experimental LEP1 data.  Using the MC implementation of our model, we
have tuned multiplicity distributions and some event shape variables
to the corresponding results as obtained from JETSET for a single
string at LEP1 energies.  To study the effect of the symmetrisation we
have then analysed and compared the reconstructed W mass of W's
generated by Pythia with and without symmetrisation included. We have
used our tuning to LEP1 energies for the symmetrised events. The
events are generated without a parton cascade, i.e. pure $\quadq$
events, in this analysis since the MC implementation of the BE-model
has not been extended to general parton configurations. We believe
that the inclusion of gluons will affect the mass reconstruction, but
it will do it in the same way whether BE symmetrisation is included or
not.

\begin{figure}[t]
\hbox{\vbox{
	\begin{center}
	\mbox{\psfig{figure=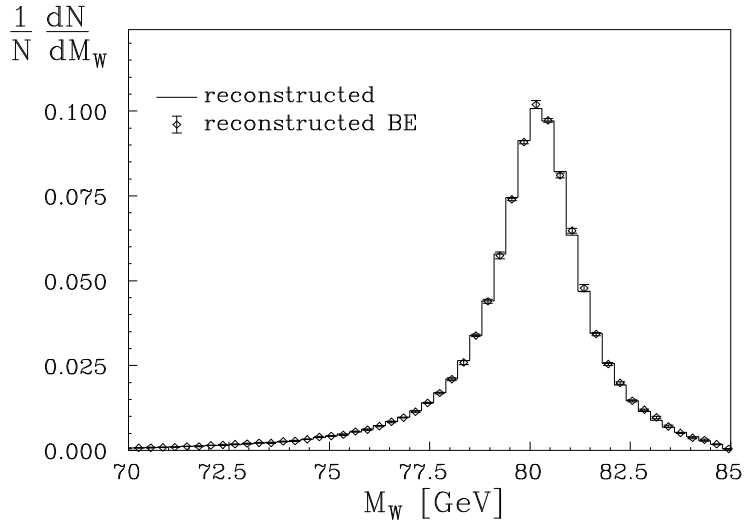,width=11cm}}
	\end{center}
}}
\caption{\em The distribution of the reconstructed W mass with (diamonds)
 and without (solid) Bose-Einstein symmetrisation turned on. The results
 are for reconstruction method~2.}
\label{f:bemass3}
\end{figure}

The use of event weights introduces statistical fluctuations which
require the generation of many events. We have generated a sufficient
number of events in order to get reasonable statistical errors for the
mass distributions. In Fig~\ref{f:bemass3} we show the reconstructed
mass for symmetrised and non-symmetrised events. As can be seen the
difference between the distributions is very small.

\begin{table}[ht]
\cebe
\begin{tabular}{|c|c|c|}
\hline
Method & $\Delta M~[MeV]$ & $\delta M~[MeV]$ \\ 
\hline
  1 & -306 $\pm$ 1 & -6 $\pm$ 8  \\
  2 & -83  $\pm$ 1 & -5 $\pm$ 7  \\
  3 & -601 $\pm$ 4 & -6 $\pm$ 6  \\
\hline
\end{tabular}
\ceen
\caption{\em Shifts in the reconstructed W masses using the different methods from Section~\ref{s:reconstruction}.  $\Delta M$ denotes the
mass shift due to the reconstruction method and $\delta M$ denotes the additional shift due to the Bose-Einstein symmetrisation.}
\label{t:be-shifts}
\end{table}
Using the same notation for the mass shifts as in
Section~\ref{sss:ciwmass} we have compiled the results for the
different reconstruction schemes in Table~\ref{t:be-shifts}.  The
mass shifts due to BE interference are all very small and compatible
with zero.  They will therefore not affect the LEP2 measurement and in
particular we conclude that the inclusion of Bose-Einstein
correlations will compared to a carefully tuned conventional
Monte-Carlo not affect the reconstruction of the W mass. It is
important to note that using event weights can in principle affect the
W mass even though we don't have any interference between the two
W's. This is however not the case with our model.


\section{Conclusions}
The previous work on the effects of BE correlations on the W mass,
with the exception of \cite{r:sharka}, are all based on the
observation that the BE effect packs identical particles closer
together. The local model \cite{r:ls95} as well as the global event
weight models \cite{r:eg96,r:beweights97} are all phenomenological
models used to estimate the influence of such a close-packing on the
masses of the two $\epq\epqa$ systems, if the two W systems
cross-talk. Our model starts from a completely different point of
view, i.e. with a quantum mechanical scenario for the particle
production dynamics, and at LEP1 energies the results obtained with
our model are in agreement with the observables on which the other
models are based. A natural consequence of our model is that we do not
expect any cross-talk due to BE effects between the W's. The
correlations between pions from different W's have been investigated by
two of the LEP experiments. The DELPHI experiment has at their present
level of statistics found no enhancement of the correlations between
pions from different W's, compared to what is expected from a pair of
uncorrelated W's \cite{r:be_delphi97} (confirmed in
\cite{r:mult_delphi97}) and ALEPH draws a similar conclusion from
their data \cite{r:be_aleph97}. Their statistics are rather poor but
if the results are confirmed when more data becomes available, it would
rule out mass shifts due to cross-talk between the two W's, in
agreement with our model.

The reconnection probability of our CI model, as in other models,
remains a free parameter. Assuming a moderate probability of 10\% the
mass shift due to CI will be very small. If we however assume a 100\%
probability the mass shift can be important, but in this case the
experimental signal of~\cite{r:gg94} should be visible. The magnitude
of the signal is a measure of the reconnection probability in our
model, and if the signal is found it can be used to estimate the
theoretical uncertainty in the mass determination.

To summarize, we conclude that neither colour nor Bose-Einstein
interference is expected to affect the W mass reconstruction at LEP2
and in particular that the theoretical uncertainties, as estimated by
our models, are much smaller than the expected experimental statistical
error.



\begin{thebibliography}{99} \vspace{-1ex}
  \bibitem{r:gg94}
	\bibl{G. Gustafson, J. H\"akkinen}{Z. Phys.}{C64}{1994}{659}
  \bibitem{r:ls95}
        \bibl{L. L\"onnblad and T. Sj\"ostrand}{Phys. Lett.}{B351}{1995}{293}
  \bibitem{r:skz94}	
        \bibl{T. Sj\"ostrand and V.A. Khoze}{Z. Phys.}{C62}{1994}{281}
  \bibitem{r:skpl94}
        \bibl{T. Sj\"ostrand and V.A. Khoze}{Phys. Rev. Lett.}{72}{1994}{22}
  \bibitem{r:eg96}
	\bibl{J. Ellis and K. Geiger}{Phys. Rev}{D54}{1996}{1967}
  \bibitem{r:beweights97}
	\bibl{S. Jadach and K. Zalewski}{Acta. Phys. Pol.}{B28}{1997}{1363} \\
	V. Kartvelishvili, R. Kvatadze and R. M{\o}ller, 
         {\it e-print, hep-ph/9704424} (1997) \\
	K. Fialkowski and R. Wit, {\it e-print, hep-ph/9709205} (1997)
  \bibitem{r:ar97}
        B. Andersson and M. Ringn\'er, {\it LU TP 97-07} and {\it hep-ph/9704383} (1997)
  \bibitem{r:JETSET}
	\bibl{T. Sj\"ostrand}{Comp. Phys. Comm.}{82}{1994}{74}
  \bibitem{r:mult_delphi97}
        DELPHI Coll., {\it Contribution to the EPS conference, Jerusalem, \\
        EPS-HEP/97-307, DELPHI 97-84 CONF 70} (1997) 
  \bibitem{r:mult_l397}
        L3 Coll., {\it Contribution to the EPS conference, Jerusalem, \\
        \mbox{\hspace{1.0cm}} EPS-HEP/97-814, L3 NOTE 2134} (1997) 
  \bibitem{r:mult_opal97}	
        OPAL Coll., {\it CERN-PPE/97-116} (1997) 
  \bibitem{r:ward97}
        D.R. Ward, {\it Talk given at the EPS conference, Jerusalem} (1997)
  \bibitem{r:yd93}
	\bibl{Yu.L. Dokshitzer, V.A. Khoze, L.H. Orr, W.J. Stirling}
		{Nucl. Phys.}{B403}{1993}{65}
	\bibl{T. Sj\"ostrand, V.A. Khoze}{Z. Physik}{C62}{1994}{281}
  \bibitem{r:ba88}
	\bibl{B. Andersson, P. Dahlqvist, G. Gustafson} 
		{Phys. Lett.}{B214}{1988}{604}
	\bibl{}{Z. Phys.}{C44}{1989}{455}
  \bibitem{r:ARIADNE}
	\bibl{L. L\"onnblad}{Comp. Phys. Comm.}{71}{1992}{15}
  \bibitem{r:ba83}
        \bibl{B. Andersson, G. Gustafson, G. Ingelman and T. Sj\"ostrand}
             {Phys. Rep.}{97}{1983}{31}
  \bibitem{r:ah86}
        \bibl{B. Andersson and W. Hofmann}{Phys. Lett.}{B169}{1986}{364}
  \bibitem{r:sharka}	
	\v{S}. Todorova-Nov\`a and J. Rame\v{s}, {\it hep-ph/9710280} (1997)	
  \bibitem{r:be_delphi97}
        \bibl{DELPHI Coll.}{Phys. Lett.}{B401}{1997}{181}
  \bibitem{r:be_aleph97}
        ALEPH Coll., {\it Contribution to the EPS conference, Jerusalem, EPS-HEP/97-590} (1997)
\end{thebibliography}
\end{document}